\documentclass[conference]{IEEEtran}

\usepackage[letterpaper, top=0.75in, bottom=1in, left=0.625in, right=0.625in]{geometry}

\usepackage{cite}
\usepackage{amsmath,amssymb,amsfonts}
\usepackage{graphicx}
\usepackage{textcomp}
\usepackage[table]{xcolor}
\usepackage{url}
\usepackage{booktabs}
\usepackage{tabularx}
\usepackage{array}
\usepackage{siunitx}
\usepackage{multirow}
\usepackage{float}
\usepackage{algorithm}
\usepackage{algorithmicx}
\usepackage{algpseudocode}
\usepackage{cuted}
\usepackage{caption}
\usepackage[hidelinks]{hyperref}

\newcolumntype{Y}{>{\raggedright\arraybackslash}X}
\definecolor{secHdr}{RGB}{245,245,245}
\definecolor{catA}{RGB}{236,244,252}
\definecolor{catB}{RGB}{253,243,232}
\definecolor{catC}{RGB}{241,236,252}
\definecolor{catD}{RGB}{235,247,242}

\definecolor{ExtSpoofC}{RGB}{0,90,200}
\definecolor{IntSpoofC}{RGB}{0,140,90}
\definecolor{ExtFloodC}{RGB}{230,140,0}
\definecolor{IntFloodC}{RGB}{0,140,90}

\makeatletter
\def\ps@headings{\def\@oddhead{\mbox{}\scriptsize\rightmark \hfil \thepage}\def\@evenhead{\scriptsize\thepage \hfil \leftmark\mbox{}}\def\@oddfoot{}\def\@evenfoot{}}
\makeatother
\IEEEoverridecommandlockouts

\def\BibTeX{{\rm B\kern-.05em{\sc i\kern-.025em b}\kern-.08em T\kern-.1667em\lower.7ex\hbox{E}\kern-.125emX}}

\newcommand{\HL}{\textit{Hop Limit }}
\newcommand{\ND}{\textit{Neighbor Discovery }}
\newcommand{\NDP}{\textit{Neighbor Discovery Protocol }}
\newcommand{\CMS}{\textit{Count-Min Sketch }}
\newcommand{\LPM}{\textsf{LPM }}

\begin{document}

\title{Rethinking IPv6 Defense: A Unified Edge-Centric Zero-Trust Data-Plane Architecture}
\author{Walid Aljoby$^{1}$, Mohammed Alzayani$^{1}$, Md.\ Kamrul Hossain$^{1}$, Khaled A.\ Harras$^{2}$\\
$^{1}$KFUPM: waleed.gobi, mohammed.zayani, g202215400@kfupm.edu.sa \\
$^{2}$CMU: kharras@cs.cmu.edu}
\maketitle

\begin{abstract}
IPv6 dependability is increasingly inseparable from IPv6 security: Neighbor Discovery (ND), Router Advertisements (RA), and ICMPv6 are essential for correct operation yet expose a broad attack surface for spoofing and flooding. Meanwhile, IPv6’s massive address space breaks per-IP reputation and makes many defenses either non-scalable or narrowly scoped (e.g., only internal threats, only RA abuse, or only volumetric floods).
We propose a \emph{zero-trust edge} architecture implemented in a single programmable data-plane pipeline that unifies four modules: external spoofing, internal spoofing, external flooding, and internal flooding. A key design choice is to enforce \emph{identity plausibility before rate plausibility}: stateless per-packet validation filters spoofed traffic early so that time-window statistics for flooding operate on credible identities.
We outline a concrete P4 design (prefix Hop-Limit bands, DAD-anchored address--port bindings, and Count-Min Sketch windowed counting) and evaluate it across a systematic 15-scenario suite spanning single-, dual-, and multi-vector compositions. We report results from a BMv2 prototype and validate the same pipeline on a Netronome NFP-4000 SmartNIC, and we discuss limitations and open directions.
\end{abstract}


\section{Introduction}
IPv6 adoption has matured from limited deployment to broad availability in enterprise, ISP, and IoT environments. Public measurements indicate a sustained rise in IPv6-enabled clients, services, and routed prefixes~\cite{w3techs_ipv6_2025,google_ipv6_2025,huston_ipv6bgp_2025}. This transition is motivated by IPv4 exhaustion and by IPv6’s ability to directly address massive device populations, including constrained IoT endpoints via 6LoWPAN/CoAP and related stacks~\cite{rolandberger_ipv6_2024}.

However, IPv6’s control-plane and local-link mechanisms also expand the attack surface. In particular, \NDP (RFC~4861) depends on ICMPv6 messages for core functions such as address configuration, router discovery, and reachability detection~\cite{narten2007rfc}. Unlike IPv4, indiscriminate ICMP filtering is not viable because ICMPv6 is operationally essential; consequently, attackers can exploit ICMPv6/\NDP to mount disruptive flooding and spoofing attacks~\cite{ripe_ipv6_security,ullrich2014ipv6,li2020towards}. These vectors are increasingly composed: modern adversaries blend spoofing (to evade attribution and diversify apparent sources) with flooding (to exhaust resources), producing traffic that can appear benign at the granularity targeted by conventional filters.

Programmable switches offer a compelling enforcement point: they enable custom parsing and match-action logic at the network edge, where mitigation is most effective. Yet existing programmable data-plane defenses often address only one dimension (spoofing \emph{or} flooding), or focus exclusively on internal threats, or require state per IPv6 address---a non-starter under IPv6’s huge and dynamic address space~\cite{liu2021jaqen,zhou2023mew,xing2021ripple,zhang2020poseidon,bai2020fastfe,zhou2024cerberus,zhang2022nethcf,li2022p4}.

\textbf{This paper proposes a unified, edge-centric zero-trust IPv6 defense architecture for programmable data planes.}
The guiding principle is strict \emph{edge verification}: do not trust source identity or traffic rates until validated by independent signals available at line rate.

\textbf{Contributions.}
\begin{itemize}
\item \textbf{Zero-trust edge architecture for IPv6.} We design a four-stage P4 pipeline that jointly mitigates \emph{external/internal spoofing} and \emph{external/internal flooding} in a single pass.
\item \textbf{IPv6-scalable identity checks.} For external sources, we validate \HL plausibility at the \emph{prefix} granularity (rather than per-host state). For internal sources, we enforce address--port binding anchored in \ND behavior, with safeguards against address-churn abuse.
\item \textbf{Data-plane flooding mitigation resilient to spoofing.} We use windowed rate tests backed by \CMS counters and intentionally place them \emph{after} spoofing filters to prevent poisoned statistics.
\item \textbf{Implementation and evaluation.} We implement the full design in P4, evaluate it on BMv2 in a heterogeneous IPv6 topology across a 15-scenario multi-vector suite, and validate the same pipeline on a Netronome NFP-4000 SmartNIC, reporting standard detection metrics.
\end{itemize}

\section{Background and Related Work}
\textbf{Spoofing and flooding in IPv6.}
Spoofing forges source identity (e.g., IPv6 address) to evade filtering or misattribute traffic. Flooding overwhelms bandwidth, CPU, state tables, or protocol processing. In IPv6, \NDP/ICMPv6 are frequent carriers: they are required for normal operation but can be weaponized for cache exhaustion, multicast amplification, or control-plane disruption~\cite{narten2007rfc,ripe_ipv6_security}.

\textbf{Programmable-switch defenses.}
A broad line of work uses P4 switches to mitigate volumetric attacks, including link-flooding and high-rate DDoS, often via sketches/counters and traffic engineering triggers~\cite{liu2021jaqen,zhou2023mew,xing2021ripple,zhang2020poseidon,bai2020fastfe,zhou2024cerberus}. Other efforts target spoofing with hop-count/TTL-based validation, which is practical in IPv4 but becomes challenging in IPv6 if it requires per-address state or assumes stable per-host hop counts~\cite{zhang2022nethcf}.

\textbf{Why IPv6 Breaks `Business-as-Usual'' Defenses.}
IPv6 adoption has matured into a safety-of-operation concern: major services increasingly observe a large share of users over IPv6, and IPv6 is foundational to IoT and modern enterprise networks. Yet three structural properties create a defense gap.

\textbf{(P1) Essential ICMPv6/ND surface.} ICMPv6 and Neighbor Discovery are required for basic operation (address resolution, reachability, SLAAC, etc.). Heavy-handed filtering can break connectivity, so defenses must be protocol-aware.

\textbf{(P2) Address-scale explosion.} Per-IP blocking/reputation is brittle in IPv6: attackers rotate addresses cheaply (temporary addresses, SLAAC), while defenders cannot store stable mappings for billions of identifiers.

\textbf{(P3) Dual adversary plane.} Dependability failures are frequently induced by both \emph{external} adversaries (Internet-side floods/spoofing) and \emph{internal} adversaries (compromised hosts abusing ND/ICMPv6, spoofing internal identities, and flooding specific services).

\textbf{IPv6-specific programmable defenses.}
Existing mechanisms address fragments. P4-NSAF defends against \emph{internal} ICMPv6 spoofing and DoS/DDoS behaviors, but does not address external adversaries~\cite{li2022p4}. RA Guard implementations focus on Router Advertisement misuse and do not provide comprehensive spoofing/flooding coverage across internal and external edges~\cite{monnich2021mitigation}. What is missing is a unified, IPv6-scale, zero-trust \emph{edge} design that remains dependable under attack composition. Our work differs by (i) adopting a unified, edge-centric zero-trust pipeline and (ii) scaling identity validation via prefix-level external checks plus dynamic internal bindings, coupled with flood mitigation robust to spoofing.

\section{Threat Model and Design Goals}
We consider an edge-deployed programmable switch that connects an internal IPv6 domain to upstream networks. The switch and its control channel are trusted. The operator can provide an initial mapping from source prefixes to plausible HL bands (or bootstrap them), and the edge can observe ND on internal ports.

\textbf{Adversary.}
We consider both \emph{external} attackers (arbitrary Internet sources) and \emph{internal} attackers (compromised hosts behind the edge). Attackers may:
(i) spoof IPv6 source addresses and manipulate \emph{Hop Limit},
(ii) flood with high-rate traffic, craft ICMPv6/\NDP packets, or stealthier bursts,
(iii) coordinate as a DDoS and vary patterns over time, and
(iv) attempt to exhaust data-plane state (e.g., churn addresses on an internal port).

\textbf{Objectives.} The defense aims to provide:
(O1) high-precision filtering of externally spoofed packets whose observed HL is implausible for the claimed source prefix;
(O2) high-precision filtering of internally spoofed packets whose source address is not bound to the ingress port;
(O3) detection and mitigation of flooding from external prefixes and internal flows within bounded detection delay; and
(O4) minimal disruption to benign IPv6 operations (low false positives).

\section{Zero-Trust P4 Defense Architecture}

\subsection{Pipeline Overview}
\begin{figure*}[!t]
  \centering
  \includegraphics[width=\linewidth]{\detokenize{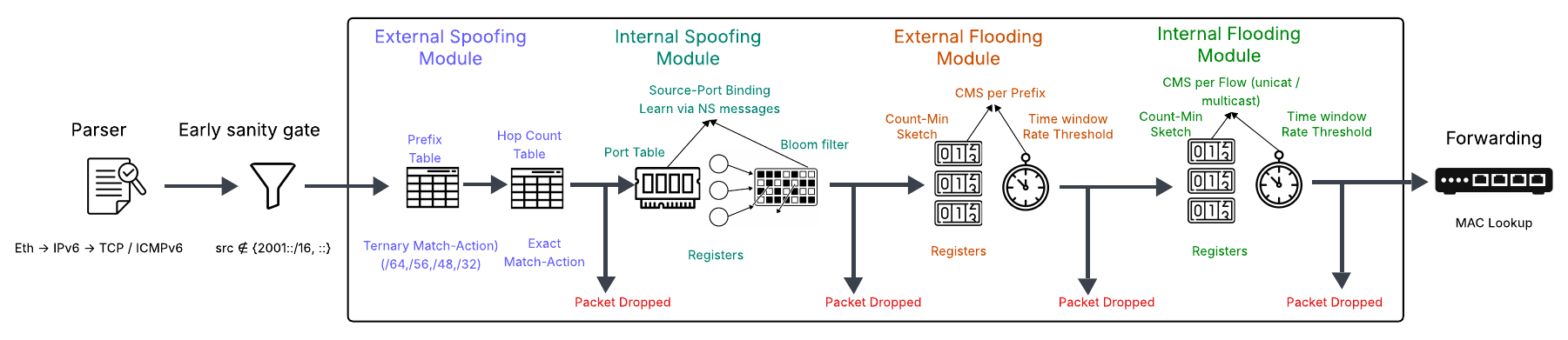}}
\caption{A unified zero-trust IPv6 defense pipeline in a P4 data plane.}
\label{fig:pipeline}
\end{figure*}

We implement four sequential modules in a single pipeline (Fig.~\ref{fig:pipeline}):
\textit{External Spoofing}, \textit{Internal Spoofing}, \textit{External Flooding}, \textit{Internal Flooding}. Spoofing checks are per-packet and stateless, while flooding checks require windowed state (sketch counters, timestamps).
\textbf{(1) External spoofing} filters packets arriving on external ports by validating (\emph{source prefix}, \HL) plausibility;
\textbf{(2) Internal spoofing} validates that internal packets’ source addresses are bound to the ingress port;
\textbf{(3) External flooding} rate-limits traffic per external source prefix using windowed counters;
\textbf{(4) Internal flooding} rate-limits internal traffic per flow identifier using windowed counters with multicast-aware thresholds.

The ordering is essential: spoofing is decided \emph{per packet} using locally available signals, enabling early removal of fake identities. Flooding requires temporal aggregation; placing it later prevents spoofed traffic from inflating counters or fragmenting apparent sources.

\subsection{External Spoofing Module: Prefix-Level HL Plausibility}
The external spoofing module filters packets whose observed IPv6 hop-limit (HL) is implausible for their claimed source \emph{prefix} at the edge. Since per-address hop-count tables are infeasible in IPv6, we operate at \emph{prefix granularity} (e.g., /32 or /48) and validate HL using longest-prefix matching.

\textbf{Metadata table and plausibility bands.}
We maintain a prefix-indexed table $\mathcal{T}$ that maps each prefix $\pi$ to an acceptable arrival-HL interval, i.e., $\mathcal{T}:\pi \rightarrow [HL_{\min}(\pi),\,HL_{\max}(\pi)]$.
For a packet arriving on an \emph{external} port, we compute $\pi \gets \LPM(a)$ for the source address $a$ and retrieve the corresponding band from $\mathcal{T}$. The packet is accepted only if its HL lies within the band for the longest-prefix match; otherwise, it is dropped. Algorithm~\ref{alg:extspoof_plaus} formalizes this check.

\textbf{Rationale.}
Attackers forging arbitrary source prefixes generally cannot match the correct arrival-HL range induced by the upstream path to the victim edge. Thus, prefix-level HL bands provide a lightweight first-line filter without per-host state.

\subsection{Internal Spoofing Module: Address--Port Binding from DAD/NS}
For packets arriving on internal ports, we enforce that each IPv6 source address is bound to a single ingress port. The goal is to prevent an internal host from spoofing another host's address on a different port.

\textbf{Learning bindings.}
IPv6 hosts perform Duplicate Address Detection (DAD) using Neighbor Discovery (ND) Neighbor Solicitation (NS) messages. We treat the first observed DAD/NS for a previously unseen address $a$ on port $p$ as a registration event and create a binding $\mathcal{M}[a]\gets p$. Thereafter, packets claiming source $a$ are accepted only if they arrive on the bound port.
To bound state growth under address rotation and to mitigate state-exhaustion attempts, we cap the number of learned addresses per port using a counter $\text{addr\_count}[p]$ and threshold $k$. If a port exceeds the cap, new bindings from that port are rejected. Algorithm~\ref{alg:intspoof_bind} formalizes the procedure.

\begin{algorithm}[!t]
\caption{External Spoofing Defense (Prefix HL Band)}
\label{alg:extspoof_plaus}
\begin{algorithmic}[1]
\Require packet $P$ with src IPv6 $a$, hop-limit $HL$, ingress port $p$
\If{$p$ is internal} \Return ACCEPT \EndIf
\State $\pi \gets \LPM(a)$
\State $[HL_{\min},HL_{\max}] \gets \mathcal{T}[\pi]$
\If{$HL \notin [HL_{\min},HL_{\max}]$} \Return DROP \EndIf
\State \Return ACCEPT
\end{algorithmic}
\end{algorithm}

\begin{algorithm}[t]
\caption{Internal Spoofing Defense (DAD + Binding)}
\label{alg:intspoof_bind}
\begin{algorithmic}[1]
\Require packet $P$ with src IPv6 $a$, ingress port $p$
\If{$p$ is external} \Return ACCEPT \EndIf
\If{$P$ is DAD/NS for $a$ and $a \notin \mathcal{M}$}
    \If{$\text{addr\_count}[p] < k$}
        \State $\mathcal{M}[a]\gets p$; $\text{addr\_count}[p] \gets \text{addr\_count}[p]+1$
        \State \Return ACCEPT
    \Else
        \State \Return DROP
    \EndIf
\EndIf
\If{$\mathcal{M}[a]=p$} \Return ACCEPT \Else \Return DROP \EndIf
\end{algorithmic}
\end{algorithm}

\begin{algorithm}[t]
\caption{External Flooding Defense (Prefix CMS)}
\label{alg:extflood_mod}
\begin{algorithmic}[1]
\Require packet $P$ with src prefix $\pi$, ingress port $p$, time $t$
\If{$p$ is internal} \Return ACCEPT \EndIf
\If{window for $\pi$ expired} reset CMS counters for $\pi$; restart window \EndIf
\State $c \gets \widehat{c}(\pi)$
\If{$c \ge \theta$} \Return DROP \EndIf
\State increment CMS counters for $\pi$
\State \Return ACCEPT
\end{algorithmic}
\end{algorithm}

\begin{algorithm}[t]
\caption{Internal Flooding Defense (Flow CMS)}
\label{alg:intflood_mod}
\begin{algorithmic}[1]
\Require flow $f$, multicast flag $m$, ingress port $p$, time $t$
\If{$p$ is external} \Return ACCEPT \EndIf
\If{window for $f$ expired} reset CMS counters for $f$; restart window \EndIf
\State $c \gets \widehat{c}(f)$
\If{$m=1$ and $c\ge \theta_m$} \Return DROP \EndIf
\If{$m=0$ and $c\ge \theta_u$} \Return DROP \EndIf
\State increment CMS counters for $f$
\State \Return ACCEPT
\end{algorithmic}
\end{algorithm}

\subsection{External Flooding Module: Windowed Rate Control per Prefix}
After spoofing filters, external flooding is mitigated by counting packets per source prefix within a time window $T_w$. Because maintaining exact per-prefix counters can be heavy at high speed, we use a Count-Min Sketch (CMS) with $d=3$ hash functions to approximate per-prefix packet counts while bounding memory.

\textbf{Prefix-keyed counting and decision rule.}
For a prefix $\pi$, the estimated count in the current window is
$\widehat{c}(\pi)=\min_{j\in\{1,2,3\}} \text{CMS}_j[h_j(\pi)]$.
For packets arriving on external ports, we (i) ensure the window for $\pi$ is current (resetting counters for $\pi$ if the window expired), (ii) compute $c \gets \widehat{c}(\pi)$, and (iii) drop if $c \ge \theta$. Otherwise, we increment the CMS counters for $\pi$ and accept, as formalized in Algorithm~\ref{alg:extflood_mod}.

\textbf{Threshold selection.} We parameterize the threshold $\theta$ from the expected service rate $r$ and the expected number of concurrently active external prefixes $n$ as
$\theta = (r/n)\cdot T_w \cdot (1+\epsilon)$,
where $\epsilon$ is an operator-chosen margin (e.g., $0.1$), yielding predictable per-prefix behavior under varying offered load.

\subsection{Internal Flooding Module:  Rate Control per Flow}
Internal flooding is often target-specific (e.g., DoS against a local server), so we track traffic at \emph{flow} granularity. The flow key $f$ can be $(\text{src},\text{dst})$ or a 5-tuple when available within parsing constraints. To bound memory and avoid maintaining explicit per-flow state, we use a CMS with $d=3$ hash functions to estimate per-flow packet counts within a time window $T_w$.

\textbf{Windowed counting and thresholds.}
For each flow key $f$, the estimated count in the current window is
$\widehat{c}(f)=\min_{j\in\{1,2,3\}} \text{CMS}_j[h_j(f)]$.
For packets arriving on internal ports, we reset counters for $f$ when its window expires and then apply separate thresholds for unicast versus multicast traffic: $\theta_u$ for unicast and $\theta_m$ for multicast. Since multicast floods can be disproportionately harmful on LANs, we configure $\theta_m < \theta_u$. Algorithm~\ref{alg:intflood_mod} formalizes the per-packet decision process.

\subsection{Implementation on BMv2 and SmartNIC}
\textbf{Targets.} In addition to BMv2 for emulation, we compiled and deployed the same P4 pipeline on a Netronome Agilio CX SmartNIC (NFP-4000) to validate behavior on hardware.
All modules execute in the ingress pipeline. Prefix extraction uses \LPM tables. Bindings and \CMS arrays use registers. Hashing uses P4-supported hash externs, producing indices into fixed-size arrays. Time windows use timestamps (BMv2 intrinsic metadata) and per-key last-seen/last-window-start registers. The control plane populates external prefix metadata; internal bindings are learned in the data plane. This separation keeps the steady-state datapath self-contained while allowing operator updates to expected prefix information. We made the implementation available online on \href{https://github.com/Muhammadkamrul/P4Defense}{GitHub}.

\textbf{External Spoofing.} We implement $\mathcal{T}$ as a \LPM-capable ternary match-action table over the IPv6 source prefix (e.g., /64,/56,/48,/32), whose action returns $(HL_{\min},HL_{\max})$ (or an expected HL) into metadata.
In ingress, we compare the packet IPv6 \texttt{hopLimit} against the returned band and set \texttt{mark\_to\_drop()} on mismatch, matching the Prefix/Hop-Count table stage in Fig.~\ref{fig:pipeline}.

\textbf{Internal Spoofing.} We learn source--port bindings in ingress using a hash-indexed register array (index = 32-bit hash(src IPv6), value = ingress\_port), updated only when parsing an ICMPv6 Neighbor Solicitation (DAD/NS) for a first-seen address.
To avoid repeated insertions, we gate learning with 3 Bloom-filter bitmaps implemented as registers (set/check via 3 hashes), and for non-NS packets we drop when \texttt{binding[src] != ingress\_port}, as shown by the Registers + Bloom-filter + Port-Table blocks in Fig.~\ref{fig:pipeline}.

\textbf{External Flooding.} We implement the 3 Count-Min Sketch arrays as three register banks, each indexed by a different hash of ($\text{src\_prefix} \,\|\, \text{prefix\_len}$), and compute $\widehat{c}(\pi)$ by reading the three counters and taking the minimum in ingress.
A window-start timestamp (from intrinsic metadata) is tracked per sketch-bank (or via double-buffered banks toggled every $T_w$), and packets are dropped when $\widehat{c}(\pi)\ge\theta$; otherwise we increment all three counters—matching the CMS-per-prefix + time-window + threshold stage in Fig.~\ref{fig:pipeline}.

\textbf{Internal Flooding.} We key the CMS by a flow identifier $f$ (e.g., $\text{src IPv6} \,\|\, \text{dst IPv6}$) and maintain 3 register-based CMS arrays updated per packet to estimate $\widehat{c}(f)$ at line rate.
We compute a multicast flag (dst is ff00::/8) and apply separate thresholds $\theta_u$ vs.\ $\theta_m$ before incrementing; exceeding the threshold triggers \texttt{mark\_to\_drop()}, matching the CMS-per-flow (unicast/multicast) + time-window + threshold stage in Fig.~\ref{fig:pipeline}.

\begin{figure*}[htbp!]
\centering
\includegraphics[width=.85\linewidth]{\detokenize{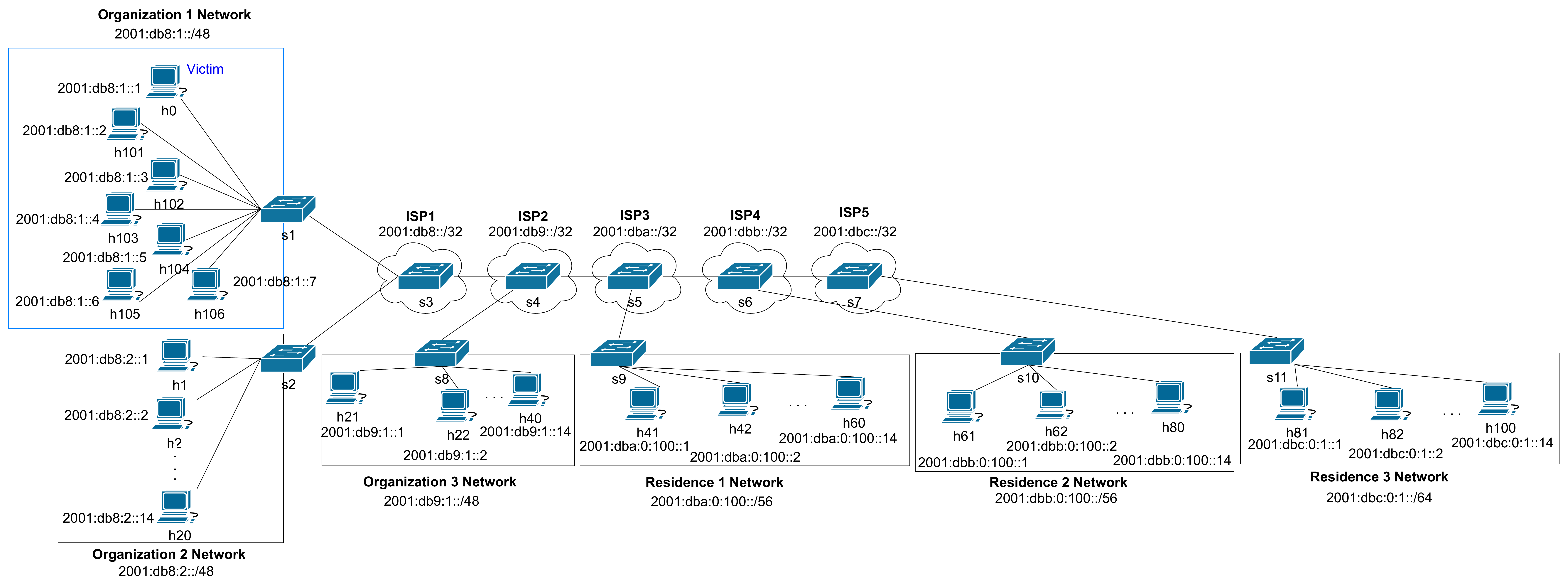}}
\caption{Evaluation setup and IPv6 topology}
\label{bsa1}
\end{figure*}

\section{Evaluation}
\subsection{Testbed Setup}
As shown in Fig.~\ref{bsa1}, we evaluate our pipeline in an emulated IPv6 topology on BMv2/Mininet that contains five ISP-level /32 prefixes and multiple downstream consumer/organization prefixes (/48,/56,/64), reflecting heterogeneous allocation practice. Our evaluation spans a fixed set of \textbf{15 attack scenarios} (Table~\ref{tab:results_p4_smartnic_all}) that systematically compose the four threat vectors in our design: \emph{internal/external spoofing} and \emph{internal/external flooding}. Specifically, scenarios 1--4 are single-vector, 5--10 are dual-vector, and 11--15 are multi-vector combinations. To complement software emulation, we compile and deploy the same P4 program on a Netronome Agilio CX 2$\times$25\,GbE SmartNIC (NFP-4000) and execute the identical 15-scenario suite; Table~\ref{tab:results_p4_smartnic_all} reports metrics for both targets side-by-side.

\textbf{Traffic generation.} Benign traffic includes \emph{ND} control messages and TCP SYN requests consistent with a web-service workload; benign \emph{ND} rates follow RFC~4861 guidance~\cite{narten2007rfc}. Attack traffic instantiates spoofing in two variants: (i) \emph{internal spoofing}, where internal adversaries inject spoofed \NDP/ICMPv6 packets while external hosts continue sending benign TCP; and (ii) \emph{external spoofing}, where external adversaries transmit spoofed TCP packets whose source IPv6 addresses are randomly selected from the external host set (h1--h100). Flooding scenarios generate high-rate traffic.

\subsection{Detection Effectiveness}
\textbf{BMv2 vs.\ SmartNIC.}
Table~\ref{tab:results_p4_smartnic_all} shows that the SmartNIC implementation consistently outperforms the BMv2 prototype across all 15 scenarios. On SmartNIC, precision is 100\% in every case and recall/accuracy remain near-perfect (recall: 98.32--100\%; accuracy: 98.48--100\%), yielding an average F1 of 99.60\%. In contrast, BMv2 exhibits lower and more variable recall (down to 83.40--86.02\% in single-vector cases), resulting in a lower average F1 of 94.26\%. This gap is expected because our defense is time-window and counter driven: SmartNIC hardware provides deterministic packet processing and stable timing, whereas BMv2’s software execution is more sensitive to host scheduling and timing jitter, which increases missed detections near window/threshold boundaries.

\begin{table}[t]
\caption{Detection performance on P4 switch and SmartNIC.}
\label{tab:results_p4_smartnic_all}
\centering
\scriptsize
\setlength{\tabcolsep}{3pt}
\renewcommand{\arraystretch}{1.12}
\begin{tabularx}{\columnwidth}{@{} Y
  S[table-format=3.2]
  S[table-format=3.2]
  S[table-format=3.2]
  S[table-format=3.2] @{}}
\toprule
\textbf{Attack Scenario} & {\textbf{Accuracy}} & {\textbf{Precision}} & {\textbf{Recall}} & {\textbf{F1}} \\
\midrule
\rowcolor{blue!10}\multicolumn{5}{@{}l}{\textbf{P4 programmable switch (BMv2)}}\\
\rowcolor{catA} 1. Internal Flooding & 96.14 & 100.00 & 85.06 & 91.92 \\
\rowcolor{catA} 2. External Flooding & 97.08 & 98.35 & 85.07 & 91.23 \\
\rowcolor{catA} 3. Internal Spoofing & 90.77 & 100.00 & 83.40 & 90.95 \\
\rowcolor{catA} 4. External Spoofing & 92.11 & 96.21 & 86.02 & 90.83 \\
\rowcolor{catB} 5. Ext.\ Flood + Ext.\ Spoof & 96.66 & 97.64 & 92.14 & 94.81 \\
\rowcolor{catB} 6. Ext.\ Flood + Int.\ Spoof & 98.10 & 99.46 & 97.77 & 98.61 \\
\rowcolor{catB} 7. Ext.\ Flood + Int.\ Flood & 94.11 & 98.60 & 88.24 & 93.13 \\
\rowcolor{catB} 8. Ext.\ Spoof + Int.\ Flood & 93.32 & 98.91 & 89.68 & 94.07 \\
\rowcolor{catB} 9. Int.\ Spoof + Int.\ Flood & 96.86 & 100.00 & 94.66 & 97.26 \\
\rowcolor{catB} 10. Int.\ Spoof + Ext.\ Spoof & 97.52 & 99.57 & 97.28 & 98.41 \\
\rowcolor{catD} 11. Int.\ Flood + Ext.\ Flood + Ext.\ Spoof & 91.00 & 88.20 & 96.11 & 91.98 \\
\rowcolor{catD} 12. Int.\ Spoof + Ext.\ Flood + Ext.\ Spoof & 95.06 & 96.13 & 98.06 & 97.09 \\
\rowcolor{catD} 13. Int.\ Flood + Int.\ Spoof + Ext.\ Flood & 90.45 & 92.65 & 94.18 & 93.41 \\
\rowcolor{catD} 14. Int.\ Flood + Int.\ Spoof + Ext.\ Spoof & 90.65 & 91.76 & 94.91 & 93.31 \\
\rowcolor{catD} 15. Full Combined Attack (All 4 Vectors) & 97.63 & 99.71 & 94.14 & 96.85 \\
\addlinespace[2pt]
\rowcolor{blue!10}\multicolumn{5}{@{}l}{\textbf{SmartNIC (Netronome Agilio CX, NFP-4000)}}\\
\rowcolor{catA} 1. Internal Flooding & 99.62 & 100.00 & 99.56 & 99.78 \\
\rowcolor{catA} 2. External Flooding & 99.29 & 100.00 & 98.72 & 99.36 \\
\rowcolor{catA} 3. Internal Spoofing & 100.00 & 100.00 & 100.00 & 100.00 \\
\rowcolor{catA} 4. External Spoofing & 98.48 & 100.00 & 98.32 & 99.15 \\
\rowcolor{catB} 5. Ext.\ Flood + Ext.\ Spoof & 99.86 & 100.00 & 99.84 & 99.92 \\
\rowcolor{catB} 6. Ext.\ Flood + Int.\ Spoof & 99.66 & 100.00 & 99.59 & 99.79 \\
\rowcolor{catB} 7. Ext.\ Flood + Int.\ Flood & 99.37 & 100.00 & 99.28 & 99.64 \\
\rowcolor{catB} 8. Ext.\ Spoof + Int.\ Flood & 98.91 & 100.00 & 98.87 & 99.43 \\
\rowcolor{catB} 9. Int.\ Spoof + Int.\ Flood & 99.71 & 100.00 & 99.68 & 99.84 \\
\rowcolor{catB} 10. Int.\ Spoof + Ext.\ Spoof & 98.88 & 100.00 & 98.82 & 99.41 \\
\rowcolor{catD} 11. Int.\ Flood + Ext.\ Flood + Ext.\ Spoof & 98.93 & 100.00 & 98.89 & 99.44 \\
\rowcolor{catD} 12. Int.\ Spoof + Ext.\ Flood + Ext.\ Spoof & 99.00 & 100.00 & 98.96 & 99.48 \\
\rowcolor{catD} 13. Int.\ Flood + Int.\ Spoof + Ext.\ Flood & 99.52 & 100.00 & 99.48 & 99.74 \\
\rowcolor{catD} 14. Int.\ Flood + Int.\ Spoof + Ext.\ Spoof & 99.12 & 100.00 & 99.09 & 99.54 \\
\rowcolor{catD} 15. Full Combined Attack (All 4 Vectors) & 99.13 & 100.00 & 99.11 & 99.55 \\
\bottomrule
\end{tabularx}
\end{table}

\section{Limitations}
\textbf{Adaptive thresholds.} Static thresholds can be brittle under diurnal and workload changes. A practical deployment can add a lightweight control-plane agent that learns baselines and updates $\theta_{\text{ext}},\theta_u,\theta_m$.

\textbf{Beyond \HL heuristics.} \HL plausibility is intentionally lightweight and topology-informed, but not cryptographic. Future extensions can incorporate additional edge signals (e.g., ingress AS hints, prefix reputation, SYN/ACK imbalance) into a multi-feature score.

\textbf{SmartNIC resource constraints.}
On the Netronome Agilio CX SmartNIC (NFP-4000), on-chip limits bound parsed headers and pipeline metadata (about $\sim$820\,B in our setup, with 4-byte alignment overhead). This motivates compact encodings; richer features may require recirculation or multi-pass designs.

\section{Conclusion}
We presented a unified, edge-centric zero-trust IPv6 defense implemented as a single programmable data-plane pipeline. By combining stateless spoofing validation (external prefix-level \HL plausibility and internal address--port binding) with stateful, windowed flooding mitigation using \emph{Count-Min Sketch}, the design remains feasible at IPv6 scale while handling multi-vector adversaries. Our BMv2 prototype and Netronome NFP-4000 SmartNIC validation show strong detection with low false positives and a clear separation between identity checks and rate enforcement, enabling practical IPv6 defenses directly in the data plane.

\bibliographystyle{IEEEtran}
\bibliography{references}

\end{document}